# Effects of temperature and strain rate on mechanical behaviors of Stone–Wales defective monolayer black phosphorene

Yan Chen[1], Hang Xiao[2,4], Yilun Liu[3] and Xi Chen[2,4,*]

[1] International Center for Applied Mechanics, State Key Laboratory for Strength and Vibration of Mechanical Structures, School of Aerospace, Xi'an Jiaotong University, Xi'an 710049, China

[2] School of Chemical Engineering, Northwest University, Xi'an 710069, China

[3] State Key Laboratory for Strength and Vibration of Mechanical Structures, School of Aerospace, Xi'an Jiaotong University, Xi'an 710049, China

[4] Columbia Nanomechanics Research Center, Department of Earth and Environmental Engineering, Columbia University, New York, NY 10027, USA

[*] E-mail: xichen@columbia.edu

**Abstract**: The mechanical behaviors of monolayer black phosphorene (MBP) are explored by molecular dynamics (MD) simulations using reactive force field. It is revealed that the temperature and strain rate have significant influence on mechanical behaviors of MBP, and they are further weakened by SW (Stone–Wales) defects. In general, the tensile strength for both of the pristine and SW defective MBP decreases with the increase of temperature or decreasing of strain rate. Surprisingly, for relatively high temperature (> 300 K) and low strain rate (< $5.0\times10^{-8}$ $fs^{-1}$), phase transition from the black phosphorene to a mixture of β-phase (β-P) and γ-phase (γ-P) is observed for the SW-2 defective MBP under armchair tension, while self-healing of the SW-2 defect is observed under zigzag tension. A deformation map of SW-2 defective MBP under armchair tension at different temperature and strain rate is established, which is useful for the design of phosphorene allotropes by strain. The results presented herein yield useful insights for designing and tuning the structure, and the mechanical and physical properties of phosphorene.

# 1. Introduction

Since graphene have been discovered as the first two dimensional (2-D) material in 2004 [1], it immediately attracted considerable interests and has been applied in many fields such as nano-sensors, composite materials, supercapacitors, catalysis, due to its excellent mechanical electronic and physical properties [2-6]. Inspired by graphene, few other 2-D materials have been discovered in succession, such as the transition metal dichalcogenide, transition metal carbides, boron nitride, etc.[7-11]

Recently, a new 2-D material has been successfully exfoliated from black phosphorus [12, 13]. Compared to semimetallic graphene, monolayer black phosphorene (MBP) has a significant advantage thanks to its inherent, direct, and appreciable band gap which can be tunable by simply controlling the layer number. Specifically, the band gap decreases from 1.51 eV to 0.59eV as the layer number increases from monolayer to five-layer [14]. MBP has a high hole field-effect mobility of 286 $cm^2/V^3s$, on/off ratio up to $10^4$ at room temperature [15], and its highest charge-carrier mobility could reach about 1,000 $cm^2V^{-1}s^{-1}$ [13]. Furthermore, MBP has several notable features such as anisotropic in-plane electrical conductance [16] and thermal conductivity [17], directional diffusion of lithium-ion [18], giant phononic anisotropy [19], and natural optical linear polarizer [20], enormously distinguished from other 2-D materials such as $MoS_2$ [15] and graphene [1].

Strain engineering can serve as a useful tool to tune the electrical and thermal properties of MBP [16, 17]. For example, negative Poisson's ratio in the out-of-plane direction have been observed [21], the preferred conducting direction of MBP can rotate by 90° with an appropriate biaxial or uniaxial strain (4−6%) in the direction parallel to the pucker [16]. Rodin et al. argued that the out-of-plane strain can significantly change the band gap and induce a semiconductor-metal transition [22]. Peng et al. found that the band gap of phosphorene experiences a direct-indirect-direct transition under both tensile and compressive strain from the relaxed structure [23]. Besides, the thermoelectric performance [19], thermal conductivity [24, 25] and photocatalyst [26] can also be tuned or tailored by strain. There is no doubt that a comprehensive theoretical analysis of the effect of strain on the mechanical behaviors of MBP is necessary for fore-mentioned electrical and thermal applications based on strain engineering.

Quite a few works have been carried out to study the mechanical behaviors of MBP. By using density functional theory (DFT) calculation, Wei et al. [27] showed that the strength of MBP is up to 18 GPa and 8 GPa in zigzag and armchair directions, respectively. The maximal in-plane Young's modulus is 166 GPa along the zigzag direction, while the minimal value is 44 GPa along the armchair direction, which exhibits strong anisotropy. Using molecular dynamics (MD) simulations, Yang et al. [28] investigated the temperature-dependent mechanical properties of MBP under biaxial and uniaxial tension as well as shear and bending deformation, and they found the elastic modulus and strength decrease with the increasing of temperature.

Like other 2-D materials, different types of defects are observed in the basal plane of MBP. The defects such as intrinsic point defects and grain boundaries in MBP are electronically inactive, in contrast to other 2-D hetero-elemental semiconductors like $MoS_2$ [29]. Hu et al. studied the stability and electronic structures of ten kinds of point defects in 2-D semiconducting phosphorene through DFT calculation [30], and they found the Stone–Wales (SW), DV-(5|8|5)-1, DV-(555|777), and DV-(4|10|4) defects have little effect on electronic properties of phosphorene, and the defective MBP is still a semiconductor with similar band gap to that of perfect MBP. Besides, the single or double vacancy in MBP exhibits lager anisotropic motion due to the low diffusion barrier compared to graphene, $MoS_2$ and BN [31]. By using first-principles calculations, Li et al. [32] have investigated the formation of line defects in single crystalline phosphorene via the migration and aggregation of point defects, including the SW, single or double vacancy (SV or DV) defects, and they found that the line defect can play an important role in modulating the electronic properties of MBP. Note that the mechanical properties as well as strain engineering of MBP can be significantly influenced by defects (which are inevitable in practice), it is quite necessary to have a comprehensive understanding of the effects of different types of defects on the mechanical behaviors of MBP.

In this work, we focus on two types of SW defects (SW-1 and SW-2) in MBP, which can be easily formed in MBP due to their low formation energy [30]. The tension behaviors of MBP with SW-1 and SW-2 defects under different temperature and strain rate are systematically studied by MD simulations using a reactive force field (ReaxFF). Phase transition and self-healing phenomena are observed for SW-2 defective MBP under armchair and zigzag tension, respectively.

## 2. Simulation model and methods

The SW defect is consisted of two 5-P rings and two 7-P rings formed by rotating one P-P bond, which is usually represented as the 55|77 defect. Considering the non-planar structure of MBP, there are two types of SW defects by rotating the two types of P-P bond in MBP, namely SW-1 and SW-2, as shown in figures 1(a) and 1(b).

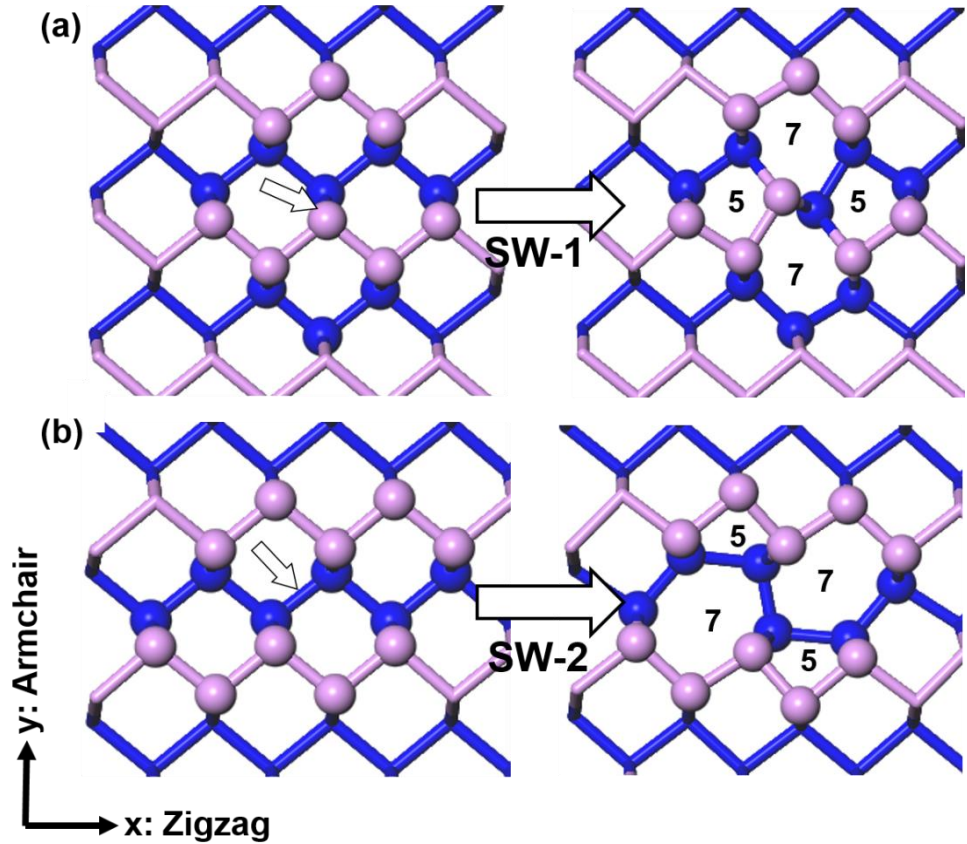

Figure 1. Schematics for 55|77(SW) defects in MBP. (a) SW-1 and (b) SW-2 defect formed by rotating one P-P bond pointed out by the arrows. Blue and purple atoms represent P atoms in lower layer and upper layer.

Periodic boundary conditions are applied in three directions of the simulation box and a vacuum layer of 3 nm is introduced in $z$ direction to eliminate the interaction of adjacent layers. The in-plane dimensions of MBP is about 10 nm × 10 nm which is large enough to eliminate the interaction of adjacent defects. The armchair direction of MBP is along $x$ axis and zigzag direction is along $y$ axis. While the thickness of MBP is set as 5.24 Å to calculate the in-plane stress in consistent with previous literatures [33, 34]. After equilibrium, uniaxial tension in armchair and zigzag direction is applied to the defective MBP by changing the dimension of the simulation box in the loading direction step by step, respectively. Time step is set to 1 fs throughout the MD simulations. During uniaxial tension, the in-plane stress perpendicular to the loading direction is zero, and system temperature is kept at 1.0 K, 100 K, 200 K, 300 K, 400 K, 500 K through Nose-Hoover barostat and thermostat, respectively. For every temperature, the strain rate varies from $10^{-6}$ fs$^{-1}$ to $10^{-8}$ fs$^{-1}$ to explore the strain rate effect on the mechanical behaviors of defective MBP. Here, the tensile strength is defined as the maximum stress during uniaxial tension and Young's modulus is defined as the slope of the stress-strain curve under tensile strain below 10%. The MD simulations were performed by using a Large-scale Atomic/Molecular Massively Parallel Simulator (LAMMPS) [35] and the interatomic interactions of MBP is described by a reactive force field of P/H systems recently parameterized by Xiao et al. [36], which can give good predictions of the mechanical, chemical and physical properties of P/H systems including defective MBP with different types of defects. Indeed, ReaxFF has a good compromise between calculation accuracy and efficiency, and has been successfully used in many systems such as polymer, hydrocarbon combustion, catalytic, etc. [37-40] Despite Stillinger-Weber potential has been used to describe the interatomic interactions of MBP [34], it usually underestimates the Young's modulus of MBP in zigzag direction and cannot correctly describe the physical and mechanical properties of MBP with SW defects and other phases of phosphorene (such as β-P and γ-P) [36]. Therefore, we choose ReaxFF as the interatomic potential to study the mechanical behaviors of defective MBP in this work.

## 3. Results and discussion

In order to verify the ReaxFF, we first calculate the SW defect formation energy $E_f$, which is defined as

$$E_f = E_{defect} - N * E_{atom}$$

where $E_{defect}$ is the potential energy of the defective MBP with $N$ atoms, and $N$ is the atom numbers. $E_{atom}$ is the energy of one atom in the pristine MBP. Compared to the previous DFT results [30, 41], the defect formation energy obtained from ReaxFF is larger, but it is in the reasonable range and better than other types of empirical potential, as shown in table 1. Besides, $E_f$ in MBP is also much smaller than that of graphene with the same defect, which indicates SW defect is easier to form in MBP [42]. Therefore, the studies of mechanical behaiors of defective MBP is of more importance.

Table 1 Defect formation energy in MBP calculated by ReaxFF, SW and DFT.

| Defect | Defect formation energy $E_f$ (eV) | | |
| --- | --- | --- | --- |
| | DFT [30] | ReaxFF | SW |
| SW-1 | 1.012 | 1.576 | 0.531 |
| SW-2 | 1.322 | 1.713 | 0 |

*3.1. Mechanical behaviors of SW defective MBP*

We first study the uniaxial tension behaviors of SW defective MBP at temperature 1.0 K and strain rate $10^{-7}$ fs$^{-1}$. The thermal and strain rate effects on the mechanical behaviors of defective MBP are presented in Section 3.2. Indeed, MBP has anisotropic mechanical properties due to its special pucker structure [27, 28], which makes MBP more flexible in armchair direction than that in zigzag direction. figure 2 shows the stress-strain curves of pristine MBP and SW defective MBP under uniaxial tensile deformation in armchair and zigzag directions, respectively. The Young's modulus of pristine MBP is 147 GPa in zigzag direction and 39 GPa in armchair direction, which is in reasonable agreement with the previous DFT results [27, 36]. Besides, the stress-strain curves of the pristine and defective MBP are almost overlapped before the fracture of the defective MBP and the effect of SW defect on the stiffness of MBP is negligible due to the very large dimensions of MBP.

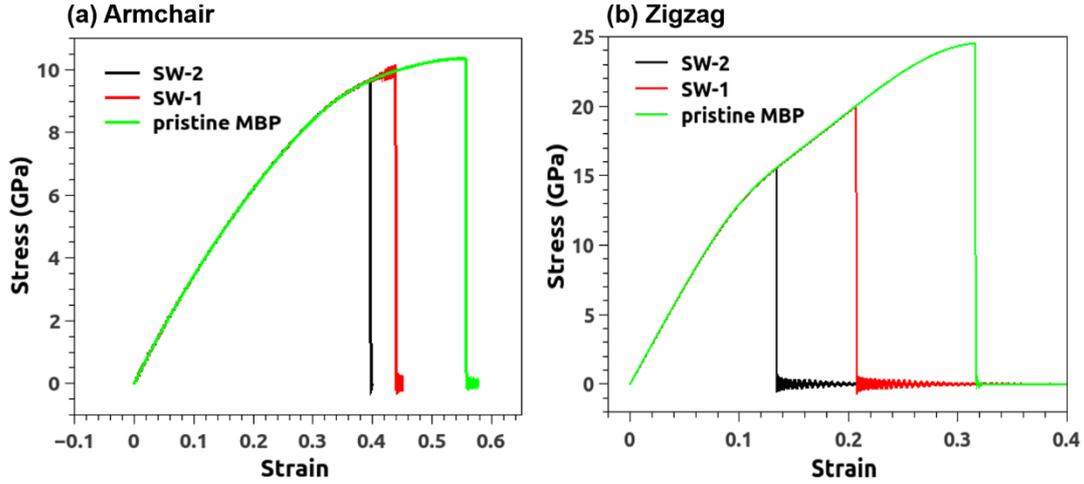

Figure 2. Tensile stress-strain curves of MBP under uniaxial tension in the armchair (a) and zigzag (b) direction for pristine MBP and MBP with SW-1 and SW-2 defect.

SW (SW-1 and SW-2) defects are formed by rotating one P-P bond in MBP, which causes large residual stress near the defects due to the distortion of phosphorene lattice, as illustrated in figure 3 and figure 4. At the armchair tensile strain of 0.0001, the P–P bond shared by heptagon–heptagon (bond 3 in figure 3(a)) in SW-1 defective MBP has tension stress and the stress can be transferred across the upper and lower layers (the upper and lower layers are connected by bond 3), as shown in figure 3(a). As the rotation angle of bond 3 is not equal to 90° and the structure of SW-1 defect is central symmetry, consequently, the stress distribution in armchair direction is also central symmetry, so that atoms C and D have the largest compressive residual stress, while atoms A and B have the largest tensile residual stress, respectively, as shown in figure 3(a). Eventually, the P-P bond with the largest tensile residual stress (bond 2, 3 in figure 3(a) corresponding to the initial fracture points in figure 3(b)) break first at strain 0.4384, reduced by 22.1% compared to the perfect MBP (fracture strain 0.5558). The strength is reduced by 3.59% compared to the perfect MBP (with strength 10.34 GPa). While for the SW-2 defective MBP, there is no symmetry between the upper and lower layers of the MBP. Therefore, there is a large residual stress concentration around the SW-2 defect, much larger than that of the SW-1 defective MBP (see the atomic stress distribution of the SW-1 and SW-2 defective MBP in figure 3(a) and 3(c)). Atoms A and B have the largest compressive residual stress, while atoms C and D have the largest tensile residual stress. However, as the tensile strain increases, the residual stress at atom A, B, C and D is partly released. While, the bond 4 and 5 have the largest tensile stress and thus fracture first at these points (as pointed out in figure 3(d) at strain 0.3958), so that the fracture strain is reduced by 28.8% and strength is reduced by 6.65% compared to the perfect MBP. In general, for both types of SW defect, the crack nucleates at the bonds with the largest tensile stress.

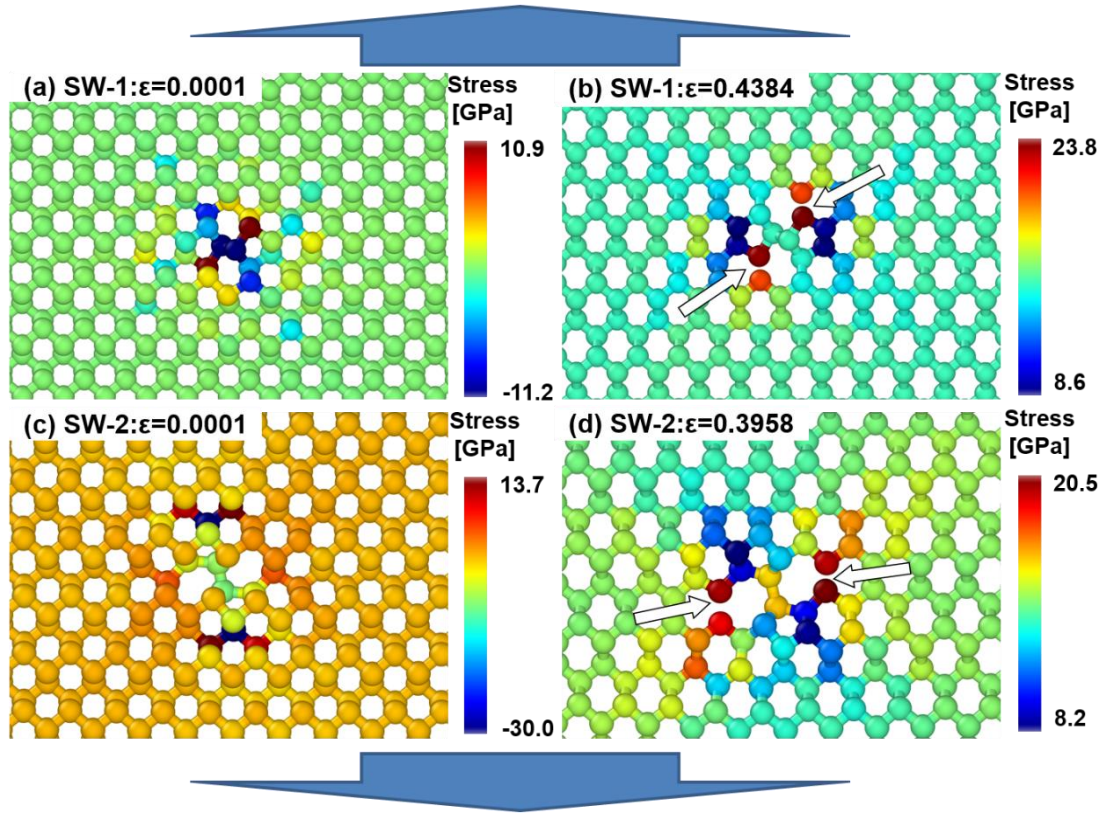

Figure 3. Atom configurations and stress distribution of SW defective MBP under armchair tension. (a) SW-1 defective MBP at strain 0.0001 and (b) at strain 0.4384. (c) SW-2 defective MBP at 0.0001 and (b) at strain 0.3958. The initial fracture points are indicated by arrows in (b) and (d). The color contour represents the atomic stress distribution in armchair direction. The numbers and capital letters indicate the bonds and atoms with large residual stress, respectively.

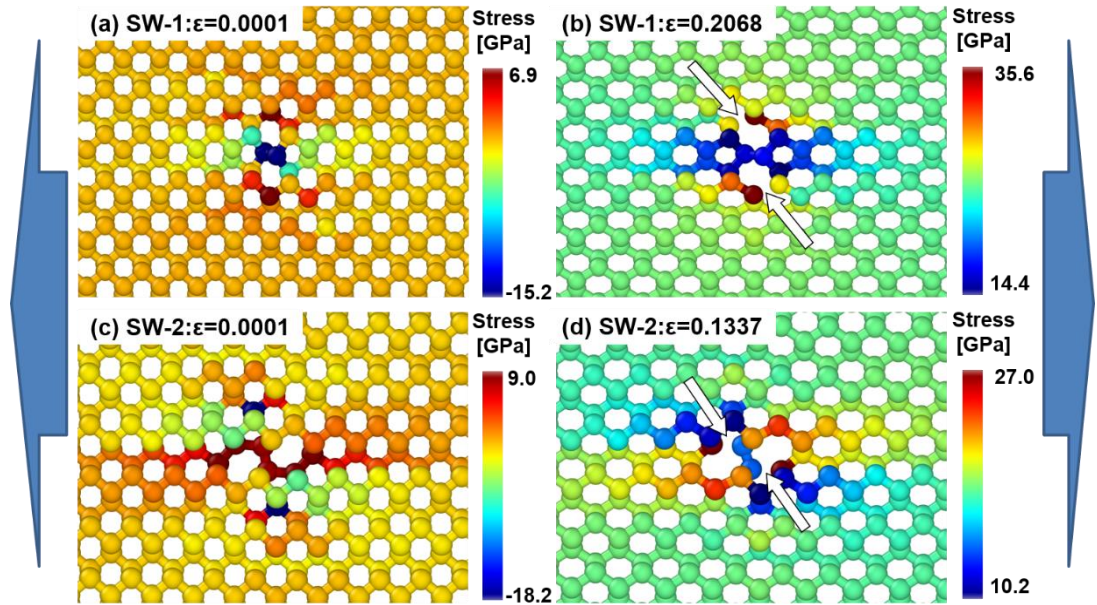

Figure 4. Atom configurations and stress distribution of SW defective MBP under zigzag tension. (a) SW-1 defective MBP at strain 0.0001 and (b) at strain 0.2068. (c) SW-2 defective MBP at strain 0.0001 and (d) at strain 0.3958. The initial fracture points are indicated by arrows in (b) and (d). The color contour represents the atomic stress distribution in zigzag direction. The numbers and capital letters indicate the bonds and atoms with large residual stress, respectively.

Comparing figure 3(b) and 3(d) for SW-1 and SW-2 defective MBP under armchair tension, both of the fractures nucleate at the P-P bond shared by the heptagon–heptagon. Meanwhile, for the former configuration the P-P bond shared by heptagon–heptagon connect the upper and lower layers, yet for the latter, it is in the same layer, so that the SW-1 defect can reproduce the pucker structure of pristine MBP to some extent which has small influence on the strength and fracture strain compared to SW-2.

Under zigzag tension, atoms C and D (figure 4(a)) in SW-1 defective MBP have the largest compressive residual stress, while atoms A and B have the largest tensile residual stress, thus bond 1 and 5 near atoms A and B break first as indicated by arrows in figure 4(b) at strain 0.2068, reduced by 34.5% compared to perfect MBP (fracture strain 0.3157). The strength is reduced by 18.5% compared perfect MBP (24.52 GPa). For SW-2 defective MBP under zigzag tension, atoms A, B, E and F have the largest tensile residual stress. As atoms A and B (the two atoms connected by the rotated bond) locate within the same layer, the SW-2 defect is asymmetrical between the upper and lower layers (figure 4(c)). The local asymmetrical tensile and compressive deformation around the defect in the upper and lower layers cause the break of the two bonds shared by pentagon-heptagon (bond 1, 3 in figure 4(c)), as pointed out in figure 4(d) at strain 0.1338. The fracture train is reduced by 57.6% and strength is reduced by 36.7% compared to the perfect MBP.

For the two types of SW defective MBP, the reduction of fracture strain and strength is more significant under zigzag tension. SW-1 defect has smaller effect on the mechanical behaviors of MBP due to its central symmetrical structure, while the asymmetrical structure of SW-2 can cause a large stress concentration on one layer, so that the reduction of strength and fracture strain is larger than that of SW-1 defective MBP. Besides, the influence of SW defects to the Young's moduli of MBP is negligible as shown in figure 2, which indicates that the influence of SW defects is "local" and hence on the fracture behaviors of MBP. Within the linear elastic deformation region, the SW defective MBP can still maintain the integrity of structure and bear the tensile load as well as the pristine MBP, whereas the fracture related behaviors are significantly reduced due to the large stress concentration near SW defects.

*3.2. Effects of temperature and strain rate*

The aforementioned fracture behaviors of SW defective MBP are obtained at very low temperature in which we aim to give some general ideas about the fracture processes of SW defective MBP. Indeed, the temperature and strain rate have an important influence on the mechanical behaviors of defective MBP, especially at high temperature. In order to investigate the temperature and strain rate effects on the mechanical properties of pristine and defective MBP, we perform MD simulations at different temperatures ranging from 1 K to 500 K and different strain rate ranging from $10^{-6}$ fs$^{-1}$ to $10^{-8}$ fs$^{-1}$.

Figure 5 illustrates the relationships between the tensile strength of pristine MBP and temperature at different strain rates. It is shown that the tensile strength of pristine MBP decreases with the increase of temperature for both of the armchair and zigzag tension in consistent with the previous literatures [43], owing to the larger thermal fluctuation at higher temperature. Besides, the pristine MBP has larger tensile strength under larger strain rate and the strain rate effect is more obvious at higher temperature, as shown in figure 5. For example, at 1.0 K the armchair and zigzag tensile strength are almost the same at different strain rates, i.e. 10.30 GPa and 24.52 GPa, respectively. Whereas, at 500.0 K the armchair and zigzag tensile strength are decreased about 5.4% and 7.7% for strain rate variation from $10^{-6}$ fs$^{-1}$ to $10^{-8}$ fs$^{-1}$, respectively. This because at a higher strain rate, MBP has shorter relaxation time to experience the thermal fluctuation and the fracture is harder to nucleate. While at lower temperature, the thermal fluctuation is also small and the strain rate effect is weakened.

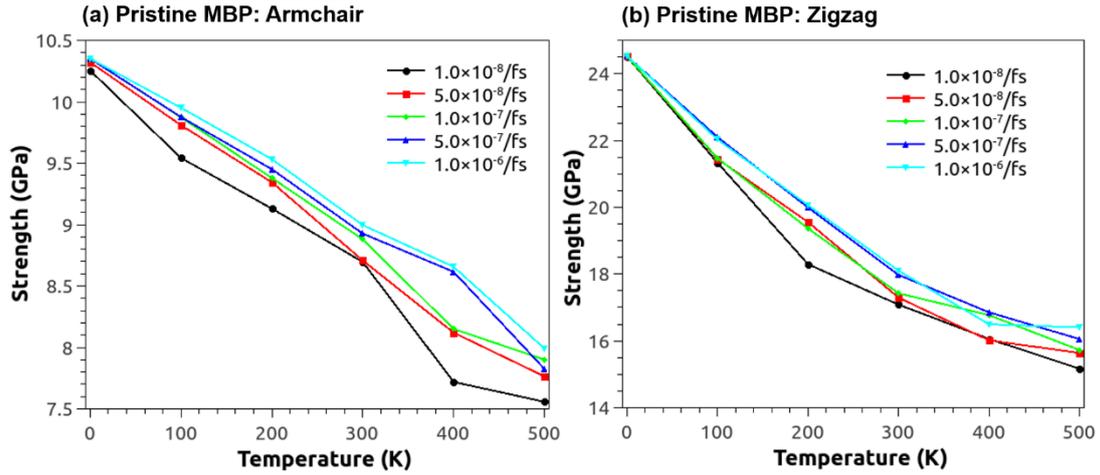

Figure 5. Relations between tensile strength of pristine MBP and temperature for different strain rate from $10^{-6}$ fs$^{-1}$ to $10^{-8}$ fs$^{-1}$ under armchair tension (a) and zigzag tension (b).

For SW-1 defective MBP, similar trends of tension strength to the temperature and strain rate can be noted, owing to the symmetrical structures, as shown in figure 6. Both of armchair and zigzag tension strength decrease with the increasing of temperature and the strain rate effect is more notable at higher temperature. The tensile strength of SW-1 MBP is smaller than that of the pristine MBP for all of the temperature and strain rate values studied herein. It is interesting that the strain rate effect of the SW-1 MBP is also more prominent than that of the pristine MBP. For example, the armchair and zigzag tension strength of SW-1 defective MBP are decreased about 12.3% and 17.5% at 500.0 K for the strain rate variation from $10^{-6}$ fs$^{-1}$ to $10^{-8}$ fs$^{-1}$ (compared to 5.4% and 7.7% for pristine MBP). Note that for SW-1 defective MBP, the fracture modes for different temperature and strain rate are all of the same, i.e. the fracture first nucleates around the P-P bond shared by heptagon–heptagon and the quickly propagates throughout the whole MBP sheet.

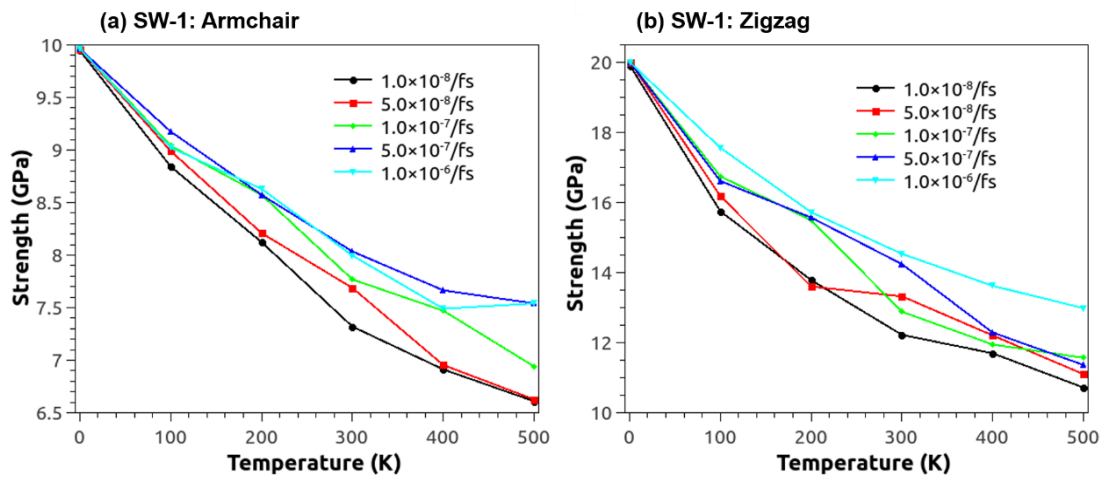

Figure 6. Relations between tensile strength of SW-1 defective MBP and temperature for different strain rate from $10^{-6}$ fs$^{-1}$ to $10^{-8}$ fs$^{-1}$ under armchair tension (a) and zigzag tension (b).

For SW-2 defective MBP, at low temperature it shares the similar tension behaviors as the pristine and SW-1 defective MBP, i.e. the tensile strength decreases with the increasing of temperature or decreasing of strain rate. However, at high temperature and low strain rate, there is an abnormal increase of the tensile strength for both of the armchair and zigzag tension, as remarked in figure 7(a) and (b). By carefully checking the atomic configurations of the SW-2 defective MBP during tension, we observe the phase transition from black phosphorene to a mixture of β-phase (β-P) and γ-phase (γ-P) under armchair tension and self-healing of the SW-2 defect under zigzag tension, which are responsible for the abnormal trend of tensile strength in figure 7(a) and 7(b).

The stress-strain curves of the SW-2 defective MBP under armchair and zigzag tension at 400 K and different strain rates are given in figure 7(c) and 7(d). The phase transition occurs for the armchair tension at 400 K and strain rate $10^{-8}$ fs$^{-1}$. The tensile stress abruptly decreases at the beginning of phase transition, followed by a long plateau stress corresponding to the propagation of β-P and γ-P, as show in figure 7(c). As the armchair tension proceeds, the tensile stress quickly increases to a peak value larger than the first peak value at the phase transition point and the fracture propagates throughout the whole SW-2 MBP at the second peak value. Because of phase transition, the fracture strain of SW-2 MBP is significantly increased. The detailed analysis of the phase transition at different temperature and strain rate is presented in Section 3.3.

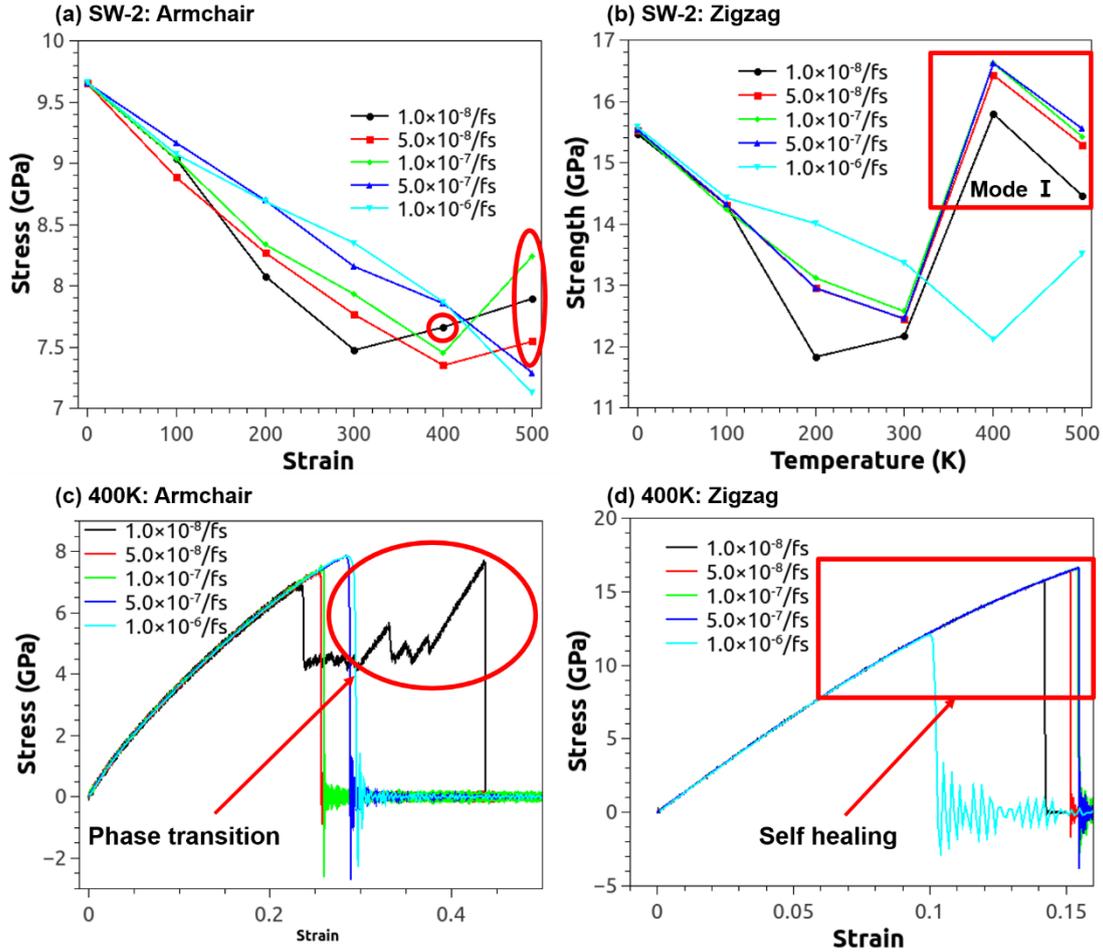

Figure 7. Relations between tensile strength of SW-2 defective MBP and temperature for different strain rate from $10^{-6}$ fs$^{-1}$ to $10^{-8}$ fs$^{-1}$ under armchair tension (a) and zigzag tension (b). The stress-strain curves of SW-2 defective MBP under armchair (c) and zigzag (d) tension at 400 K and different strain rate from $10^{-6}$ fs$^{-1}$ to $10^{-8}$ fs$^{-1}$. The marked region in (a) and (c) corresponds to the phase transition (Mode C in section 3.3) and the marked regions in (b) and (d) correspond to the self-healing of SW-2 defect.

The self-healing of the SW-2 defect occurs for the zigzag tension at 400 K and strain rate smaller than $10^{-6}$ fs$^{-1}$. From the atomic configurations of the SW-2 defective MBP under zigzag tension (figure 8), it is shown that fracture nucleates near SW-2 defect at zigzag tension 0.101 for strain rate $10^{-6}$ fs$^{-1}$, while for strain rate $10^{-8}$ fs$^{-1}$, the SW-2 defect disappears at zigzag tension strain 0.0633 and the SW-2 defective MBP becomes a perfect MBP, as shown in figure 8(b). Because of the self-healing of SW-2 defect, the tensile strength of the SW-2 defective MBP under zigzag tension at 400 K and strain rate smaller than $10^{-6}$ fs$^{-1}$, are significantly increased, as shown in figure 7(b) and 7(d). Indeed, the atomic structures of SW-2 are not symmetrical at the upper and lower layer of the MBP, and local stress concentration cause the P-P bond around the 5-P rings and 7-P rings are prone to rotate during tension so as to activate the phase transition and self-healing. Apart from the phase transition and self-healing, the mechanical behaviors of SW-2 defective MBP are similar to that of pristine MBP and SW-1 defective MBP, i.e. the tensile strength decreases as the increasing of temperature or decreasing of strain rate and the strain rate effect is more significant at higher temperature, which is also consistent with the mechanical behaviors of polymer [44] and graphene [45].

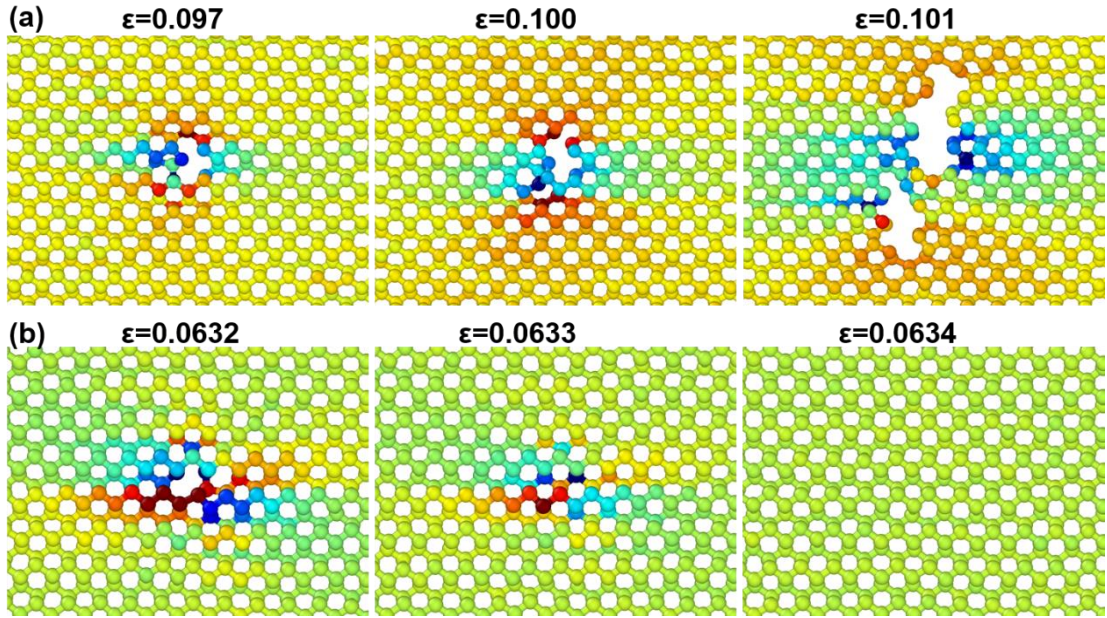

Figure 8. Atomic configurations and stress distribution for SW-2 defective MBP under zigzag tension at 400 K and different strain rate of $10^{-6}$ fs$^{-1}$ (a) and $10^{-8}$ fs$^{-1}$ (b). The self-healing of the SW-2 defect is observed at zigzag tension strain 0.0633 in (b). The color contour represents the atomic stress distribution in zigzag direction.

*3.3. Phase transition of MBP with SW-2 defect*

Recently, different phases of phosphorus element has been discovered in experiment [46, 47] and DFT calculation [48-50]. However, the phase transition of the phosphorus elements caused by mechanical loading has not yet been reported. Through our MD simulations, we have found the phase transition from black phosphorene to a mixture of β-P and γ-P for the SW-2 defective MBP under armchair tension. The phase transition is sensitive to the temperature and strain rate. In this section, we will analyze the phase transition process, the temperature and strain rate effects on the phase transition in detail.

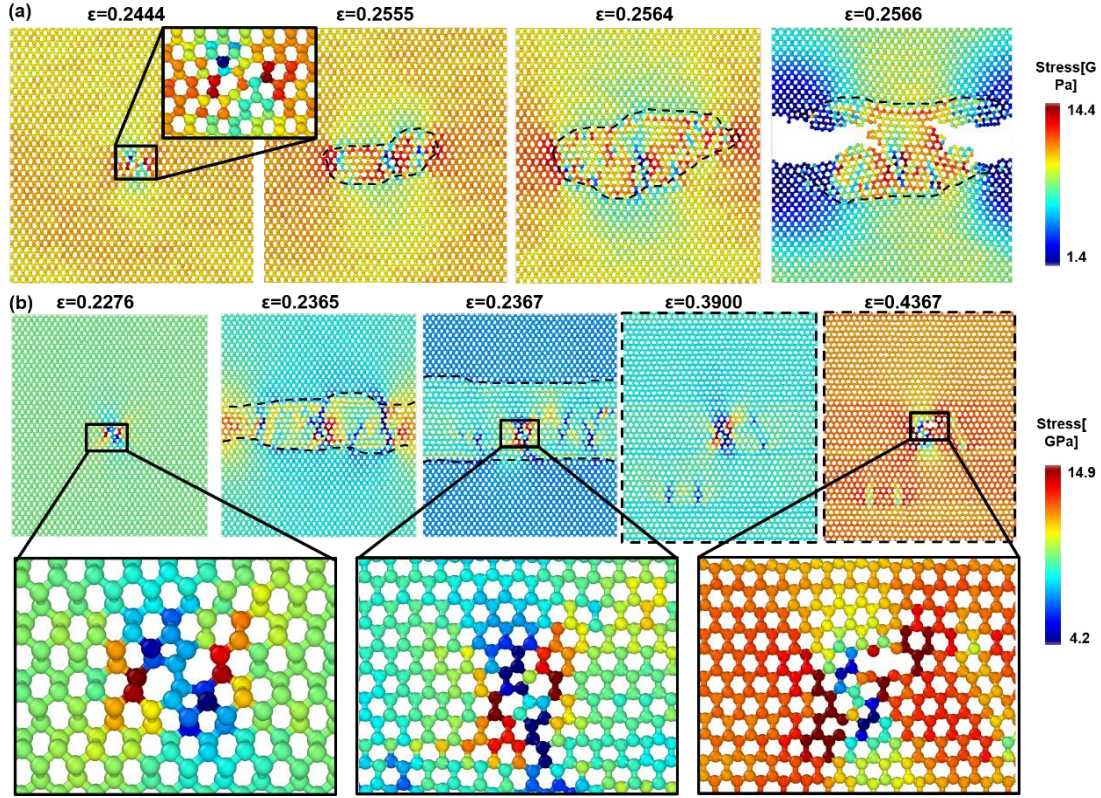

Figure 9. Atomic configurations and stress distribution for SW-2 defective MBP under armchair tension at 400K and different strain rate of $5\times10^{-8}$ fs$^{-1}$ (a) and $10^{-8}$ fs$^{-1}$ (b). Phase transition regions are marked by black dashed line. The inset in (a) and lower figures in (b) show the atomic configurations near the SW-2 defect. Color contour represents the atomic stress distribution in armchair direction.

Figure 9 illustrates the atomic configurations and stress distribution of SW-2 defective MBP under armchair tension at 400K and different strain rate of $5\times10^{-8}$ fs$^{-1}$ and $10^{-8}$ fs$^{-1}$. It is observed that the phase transition mode is different for different strain rates. For the strain rate of $5.0\times10^{-8}$ fs$^{-1}$, the phase transition initiates near the SW-2 defect at armchair strain 0.2444, and then propagates quickly to form a small phase transition region around the SW-2 defect as marked by dashed line in figure 9(a). Finally, the SW-2 defective MBP fracture at tension strain 0.2566 before the phase transition region spreads to the whole sheet. For this phase transition mode (denoted as Mode B), the crack nucleates at the boundary of the phase transition region due to the stress concentration at the boundary. Whereas for the strain rate of $10^{-8}$ fs$^{-1}$, the phase transition initiates from the SW-2 defect at tension strain 0.2276, and then quickly spread across the MBP sheet to form a phase transition band at strain 0.2367. Afterwards, the area of the phase transition region continuously increases, until the whole MBP become a mixture of β-P and γ-P at tension strain 0.3900, as shown in figure 9(b). Finally, the fracture nucleates at the SW-2 defect upon strain 0.4367 (denoted as Mode C) and the fracture strain is significantly increased.

The phase transition pattern for the case shown in figure 9(b) is given in figure 10(d). In order to clarify the black phosphorene, β-P and γ-P, they are represented as purple, red and blue color in figure 10(d) and the atomic configurations of black phosphorene, β-P and γ-P are given in figures 10(a)-(c). From figure 10(d), it is observed that the phase transition originates from the SW-2 defect and the black phosphorene first changes to β-P. Note that the P-P bond 1 (see inset of figure 10(d)) is subjected to a localized torque due to the asymmetrical structure of SW-2 defect. When the critical tension strain is reached, i.e. 0.2276 for 400 K and strain rate $10^{-8}$ fs$^{-1}$, the P-P bond 1 flips along the axis a-a (see inset of figure 10(d)) that locates in MBP plane and is normal to the P-P bond 1, so as to initiate the phase transition from black phosphorene to β-P. Large thermal fluctuation at higher temperature and sufficient relaxation time are beneficial for the SW-2 defective MBP to overcome the phase transition energy barrier. Afterwards, the phase transition region spreads quickly across the MBP sheet in a cascading manner, and the phase transition region is consisted of β-P and γ-P (see the third instant in figure 10(d)). With the armchair tension strain increasing, the β-P and γ-P in the phase transition region prefer to form bands cross the MBP in the zigzag direction (see the forth to seventh events in figure 10(d)). At tension strain 0.3900, the whole SW-2 defective MBP sheet becomes a mixture of β-P and γ-P. The systematical studies of the transition between β-P and γ-P, the control of β-P to γ-P ratio, and the electrical and thermal properties of the β-P and γ-P mixture will be presented in our future works.

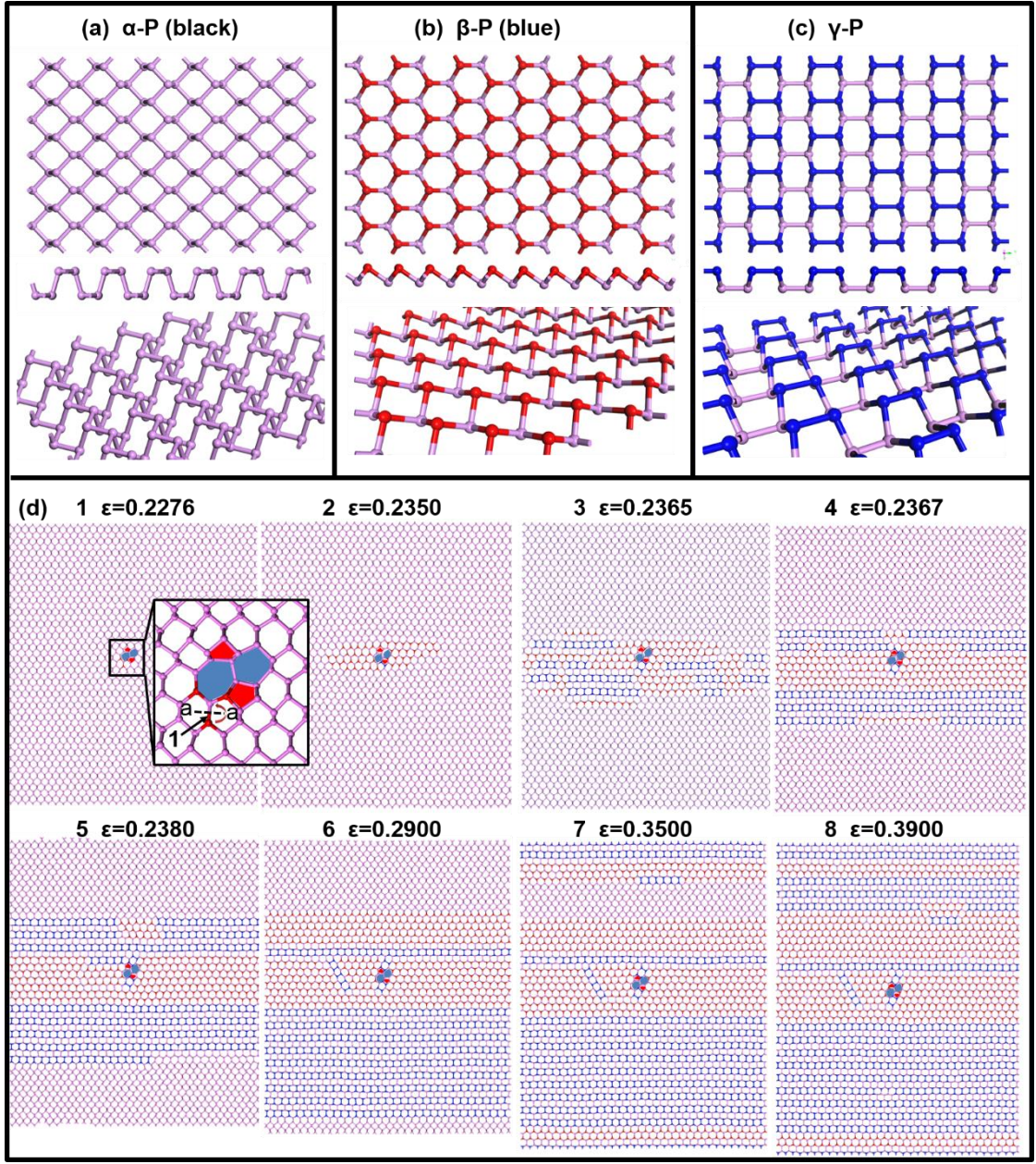

Figure 10. The atomic configurations of black phosphorene (a), β-P (b) and γ-P (c). (d) Phase transition pattern of SW-2 defective MBP under tension in the armchair direction at 400K and strain rate $10^{-8}$/fs corresponding to the phase transition in Figure 9 (b). Purple, red and blue color in (d) represent black phosphorene, β-P and γ-P. The inset in (d) shows the atomic structure of SW-2 defect, and the pentagons are represented as red color and two heptagons are represented as blue color.

The phase transition of the SW-2 defective MBP under armchair tension is sensitive to temperature and loading rate. In general, high temperature and low strain rate are beneficial to the phase transition. Based on phase transition regions and fracture nucleation patterns, we can divide the phase transition phenomena of the SW-2 defective MBP under armchair tension into three modes, as shown in figure 11. Mode A occurs at low temperature and the phase transition is not observed until the fracture of the whole MBP sheet. For this mode, the fracture first nucleates at the SW-2 defect. Mode B occurs at medium temperature and partial phase transition of MBP is observed. The fracture first nucleates at the boundary of the phase transition region due to large residual stress, as shown in figure 9(a). Mode C occurs at high temperature and low strain rate and the whole MBP become a mixed phase of β-P to γ-P. For this mode, the fracture also nucleates at the SW-2 defect and the fracture strain is significantly increased, as shown in figure 9(b).

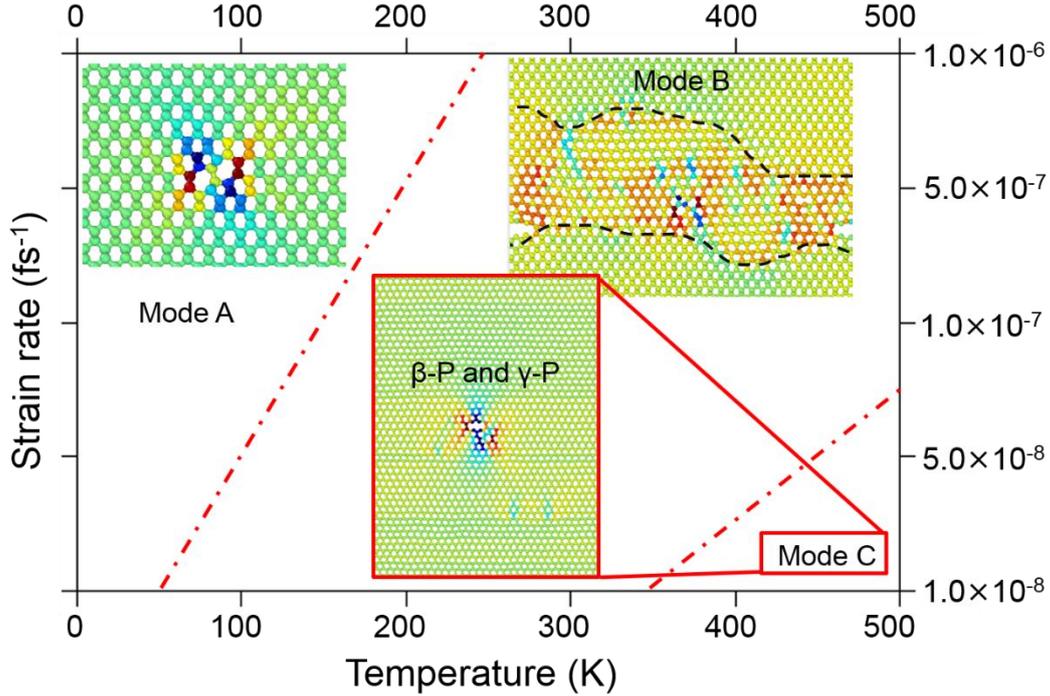

Figure 11. The patterns of phase transition of SW-2 defective MBP under armchair tension at different temperature and strain rate. Mode A represents no phase transition and fracture nucleates at the SW-2 defect; Mode B represents partial phase transition and fracture nucleates at the phase transition boundary; Mode C represents phase transition of the whole MBP and fracture nucleates at the SW-2 defect.

In order to check the stability of the phase transition, two cases of the SW-2 defective MBP under armchair tension (i.e. armchair strain 0.2952 at 400 K and $5\times10^{-8}$ fs$^{-1}$ corresponding to Mode B; armchair strain 0.3900 at 400 K and $1\times10^{-8}$ fs$^{-1}$ corresponding to Mode C) are unloaded to zero tension stress and equilibrated through NPT with Nose-Hoover thermostat and barostat at 400 K for 200 ps. The deformed and relaxed configurations are given in figure 12. For Mode B, almost all phase transition regions are kept after unloading and equilibrium, except a few regions near the phase transition boundary due to large residual stress at the boundary, as shown in figure 12(a). While, for Mode C, all phase transition regions are kept, as shown in figure 12(b). Because of the retained β-P to γ-P, there is some residual tension strain after the unloading to zero stress, especially for Mode C. As the mismatch of the lattice among black phosphorene, β-P and γ-P, buckled configures are observed for the relaxed configurations and the MBP for Mode B has larger curvature due to the larger grain boundary and unregular phase transition region.

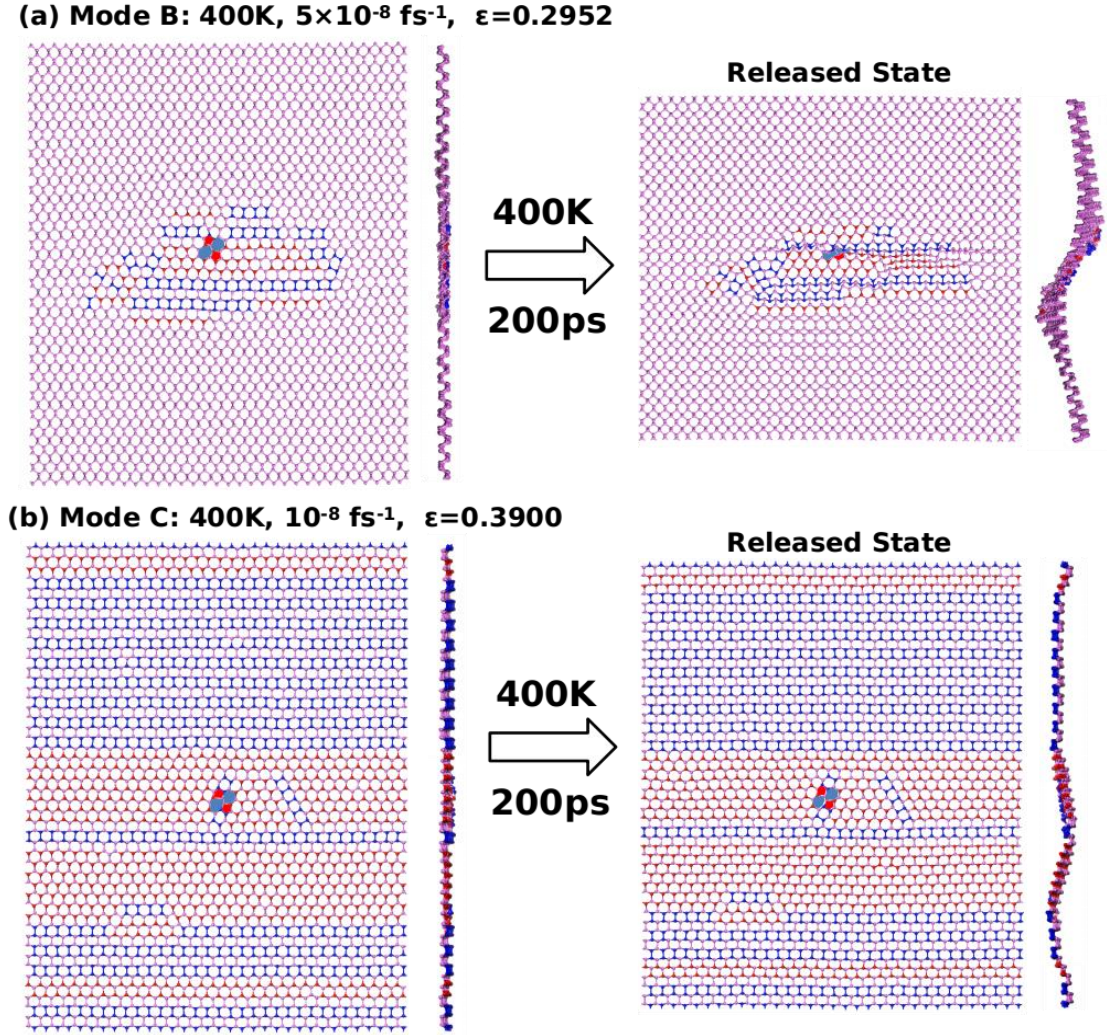

Figure 12. Relaxed atomic configurations of the phase transition region after unloading for Mode B (a) and Mode C (b). Purple, red and blue color represent black phosphorene, β-P and γ-P.

## 4. Conclusions

The mechanical behaviors of the SW (SW-1 and SW-2) defective MBP are systematically studied through MD simulations using ReaxFF. The tensile strength and fracture strain of MBP are weakened by SW defect for both of the armchair and zigzag tension, while the Young's modulus is almost unaffected. In general, the tensile strength and fracture strain of the SW defective MBP decreases as the increasing of temperature or decreasing of strain rate, and the strain rate effect is more significant at higher temperature, which is similar with the pristine MBP. However, phase transition from black phosphorene to a mixture of β-P to γ-P is observed for SW-2 defective MBP under armchair tension at high temperature and low strain rate, while self-healing of the SW-2 defect is observed for zigzag tension. Because of the phase transition and self-healing, the tensile strength and fracture strain are increased. The diagram of the phase transition of SW-2 defective MBP under armchair tension at different temperature and strain rate is given, which may provide a mechanical method to synthesis β-P and γ-P. Overall, the results presented herein are beneficial for the potential applications of MBP in electronic, thermal and optical devices with defect and mechanical loading.

## Acknowledgment

Y.L. acknowledges the support from the National Natural Science Foundation of China (No. 11572239) and National Key Research and Development Program of China (No. 2016YFB0700300). X.C. acknowledges the support from the National Natural Science Foundation of China (Nos. 11372241 and 11572238), ARPA-E (DE-AR0000396) and AFOSR (FA9550-12-1-0159).